\documentclass[twocolumn,prl,showpacs,amsmath,amssymb]{revtex4}

\usepackage{graphicx}
\usepackage[latin1]{inputenc}
\usepackage{dcolumn}
\usepackage{bm}
\usepackage{epsfig}

\begin{document}

\preprint{APS/123-QED}

\title{Correlation dimension of complex networks}

\author{Lucas Lacasa$^{1}$}
\email{lucas.lacasa@upm.es}
\affiliation{$^1$ Departamento de Matem\'{a}tica Aplicada y Estad\'{i}stica,\\ ETSI Aeron\'{a}uticos, Universidad Polit\'{e}cnica de Madrid, Spain\\
$^2$ Departamento de F\'{i}sica de la Materia Condensada, Universidad de Zaragoza, Spain\\
$^3$ Institute for Biocomputation and Physics of Complex Systems (BIFI), Universidad de Zaragoza, Spain}%

\author{Jes\'{u}s G\'{o}mez-Garde\~{n}es$^{2,3}$}
\affiliation{$^1$ Departamento de Matem\'{a}tica Aplicada y Estad\'{i}stica,\\ ETSI Aeron\'{a}uticos, Universidad Polit\'{e}cnica de Madrid, Spain\\
$^2$ Departamento de F\'{i}sica de la Materia Condensada, Universidad de Zaragoza, Spain\\
$^3$ Institute for Biocomputation and Physics of Complex Systems (BIFI), Universidad de Zaragoza, Spain}%

\date{\today}

\begin{abstract}
We propose a new measure to characterize the dimension of complex networks based on the ergodic theory of dynamical systems. This measure is derived from the correlation sum of a trajectory generated by a random walker navigating the network, and extends the classical Grassberger-Procaccia algorithm to the context of complex networks. The method is validated with reliable results for both synthetic networks and real-world networks such as the world air-transportation network or urban networks, and provides a computationally fast way for estimating the dimensionality of networks which only relies on the local information provided by the walkers.\\
\textbf{Preprint version merging the main article and the supplementary material. To be published in Physical Review Letters.}
\end{abstract}

\pacs{05.45.Ac,05.45.-a,89.75.Fb}

\maketitle

Network science has influenced the recent progress in many areas of statistical and nonlinear physics \cite{barabasi}. The discovery of the real architecture of interactions of many systems studied under the former disciplines \cite{rev:albert,rev:newman,rev:bocc} changed the usual mean-field way to tackle problems arising in sociology, biology, epidemiology and technology among others \cite{vespignani}. Furthermore, the blossom of the network theoretical machinery \cite{newman_book}, has provided a forefront framework to interpret the relations encoded in large datasets of diverse nature and fostered the application of new techniques, such as community detection algorithms \cite{santo}, to coarse-grain the complex and hierarchical landscape of interactions of real-world systems.


Recently, geometrical concepts have been exploited to describe and classify the structure of complex networks beyond purely topological aspects \cite{emilio,boxcovering, serrano,serrano1}.
In particular, the box-counting technique, widely used for estimating the capacity dimension $D_0$ of an object, has been recently extended, as a box-covering algorithm, to characterize the dimensionality of complex networks \cite{boxcovering, boxcovering2, boxcovering3, jstat}. This technique proceeds by calculating the number $N$ of boxes of Euclidean volume $L^d$ required to cover an object, being the capacity dimension  $D_0$ of such object given by $D_0=\lim_{L\rightarrow 0} \frac{\log N}{\log(1/L)}$. The capacity dimension $D_0$ is thus seen as an upper bound to the Hausdorff dimension.

The box-covering approach, while being the most natural and elegant extension of the concept of fractal dimension to networks, suffers from some difficulties. First, in order to tile the network and to unambiguosly relate the box-covering and capacity dimensions, the object under study must be embedded in a metric space, something that does not apply in the more general case of a complex network.
This subtle problem can be overcome by restricting to spatially embedded complex networks \cite{boxcovering3}. A second important issue is the need of full knowledge of the network topology in order to perform the box-covering procedure. This constraint faces the limitations related to storing the complete network backbone, indeed, the computation of the capacity dimension becomes unpractical for embedding dimensions larger than $3$ \cite{review}. Finally, another related problem is that of finding the optimum covering, whose computational complexity is NP-hard \cite{jstat}.

The above difficulties for calculating the capacity dimension of a self-similar object are however circumvented
in the dynamical systems literature by, instead, calculating its correlation dimension \cite{review}. Here we take advantage of this alternative characterization to compute the dimension of complex networks, relying on an extension of the Grassberger-Procaccia algorithm \cite{seminal, review}.  The key idea to extend this concept to the network realm is to generate random walkers surfing the network whose dimension we want to estimate and to study their actual trajectories as time series. As a byproduct, the extension of this technique opens the door to the use of the theoretical machinery inherited from the ergodic theory of dynamical systems in the characterization of the structure of networks.

\begin{figure*}
\centering
\includegraphics[width=1.7\columnwidth]{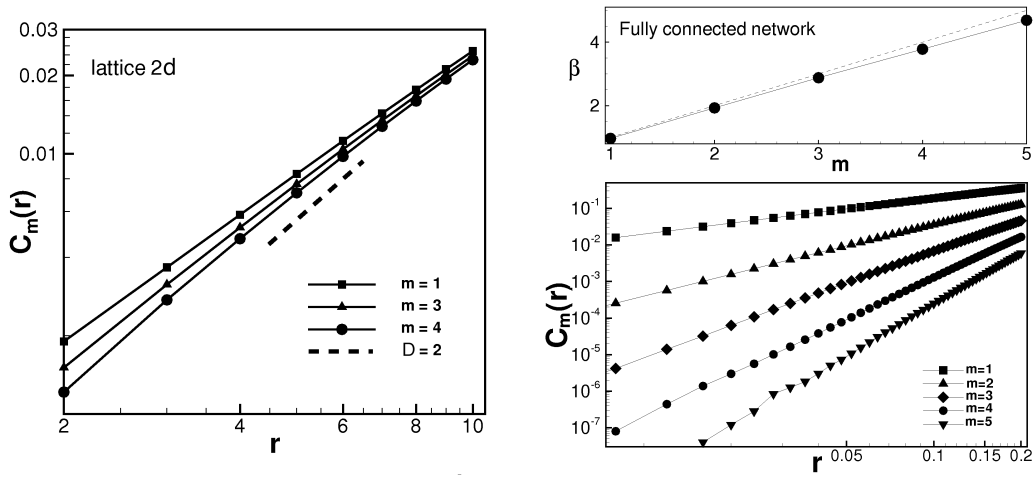}
\caption{(Left) Log-log plot of the correlation sum $C_m(r)$ as a function of similarity $r$, for a series of $4\cdot10^4$ data extracted from an unbiased random walker in a 2d lattice of $1000$ nodes (with correlation dimension $2$) where ${\bf v}=(v_x,v_y)$ and $v_x,v_y \in [1,1000]$, for different embedding dimensions $m$. There exists a scaling regime where the slope of the correlation sum approaches $2$ for increasing values of $m$ (for $m=4$, we find $\beta=1.92\pm0.1$). (Right panel, bottom) Log-log plot of the correlation sum $C_m(r)$ as a function of similarity $r$, for a series of $10^4$ data generated by a random walker over a fully connected network, for different values of the embedding dimension $m$. In this network each node is labeled with a real value ${\bf v}\equiv v \in [0,1]$. In all cases a scaling $C_m(r)\sim r^{\beta_m}$ is found. (Right panel, top) Correlation exponent $\beta_m$ as a function of the embedding dimension $m$. The correlation exponent increases linearly with the embedding dimension $m$, what suggests an infinite-dimensional network.}
\label{validation}
\end{figure*}

Indeed, the study of the structure of networks relying on the theory of stochastic processes, such as random walks, has been successfully applied in the past for designing ranking algorithms, such as Google \cite{pr}, and unveiling the community structure \cite{rosvall} or the nature of degree-degree correlations \cite{vito} in complex networks. In our case, although random walkers are stochastic processes which have an underlying infinite-dimensional attractor, their trajectories are expected to evidence temporal correlations intimately related to the structure of the underlying network that confines their movement. Thus, in the case of self-similar correlations an associated dimension can be properly defined, yielding a reliable \cite{note1} estimation of the underlying network's dimension. In the rest of the paper, after introducing the method, we present some results for both synthetic and real spatial networks \cite{barthelemy} and compare with those results obtained by means of box-covering techniques.\\

We start by introducing the method for estimating the correlation dimension in complex networks. Let ${\cal G}$ be an undirected network with $N$ nodes and $L$ links so that each node $i$ of ${\cal G}$ is labeled with a generic vector $\textbf{v}_i$, where $\textbf{v} \in \mathbb{R}^d, \  \text{or} \in \mathbb{N}^d$ when the space is discrete. Consider a trajectory of length $n$ generated by an ergodic random walker surfing the network ${\cal G}$, described by
the series $\{{\bf v}_1,{\bf v}_2,\dots,{\bf v}_n\}$. Note that in the case of spatially embedded networks, $\textbf{v}_i$ uniquely characterizes the position of node $i$ in the underlying space. For instance, in a 2-dimensional space, $\textbf{v}_i=(v_x,v_y)^T$ and the series reads $\{v_x(1),v_y(1),v_x(2),v_y(2),\dots,v_x(n),v_y(n)\}$.
This series is the object of study in order to describe the geometry and dimension of ${\cal G}$ \cite{note2} and the first step is to apply embedding techniques to $\{\textbf{v}_t\}_{t=1}^n$. Inspired by Taken's embedding theorem \cite{takens1}, we proceed to construct the surrogate vector-valued series $\{{\bf V}(t)\}$ where ${\bf V}(i) \in \mathbb{R}^{m\cdot d}$:
\begin{equation}
{\bf V}(i)=[\textbf{v}_{i+1},\dots,\textbf{v}_{i+m-1}],
\end{equation}
where $m$ is the so called embedding dimension.
Then, the correlation sum function $C_m(r)$ is defined as the fraction of pairs of vectors whose distance is smaller than some similarity scalar $r \in \mathbb{R}$ \cite{note3}:
\begin{equation}
C_m(r)=\frac{2\sum_{i<j} {\theta}(\|{\bf V}(i) - {\bf V}(j)\|-r)}{(n-m)(n-m+1)}\;,
\end{equation}
where $\theta(x)$ is the Heaviside step function, and $\|\cdot\|$ is usually a p-norm $\|{\bf x}\|_p= \bigg[\sum_i |x_i|^p\bigg]^{1/p}$.
Without loss of generality, here we choose $\|\cdot\|$ as the $L^\infty$ norm, $\|{\bf x}\|_\infty= \max( |x_1|,|x_2|,\dots,|x_n| )$, that induces the so called Chebyshev distance. Note that the use of the Euclidean norm was originally proposed in \cite{seminal}, while the use of maximum norm was used by Takens in \cite{takens}.

Based on arguments from ergodic theory \cite{seminal, review}, we conjecture that when the series is extracted from the trajectory of a walker surfing a network with well defined dimension, for sufficiently long series and sufficiently small values of $r$,
$C_m(r)$ evidences a scaling regime such that:
\begin{equation}
\lim_{r\rightarrow0}\lim_{n\rightarrow\infty}\frac{\log(C_m(r))}{\log(r)}=\beta_m\;,
\end{equation}
where $\beta_m\rightarrow \beta$ for sufficiently large embedding dimension (whereas Whitney's theorem provides as an upper bound $m>2D+1$, in this case we will show that the correlation dimension saturates for relatively small values of $m$ as the random walker series are noise-free). Thus, $\beta$ is the estimate of the correlation dimension of the underlying space, here the complex network under study. Note that, in practice, the scaling regime is expected to appear only at an intermediate range $r_0<r<r_1$, where $r_0\sim {\cal O}(10^{-2}\langle v_x \rangle)$ is a lower cut-off due to poor statistics (noise regime) whereas $r_1$ is an upper cut-off due to nonlinear effects (macroscopic regime) \cite{review, book}. We will show that the walker size required to capture the network's geometry is only of order ${\cal O}(n)$, where $n$ is the number of nodes.

In order to validate the above method, we first address a synthetic network that can be understood as a discrete limit of a smooth metric space with a well defined Hausdorff dimension. In the left panel of Fig. \ref{validation} we plot the correlation function $C_m(r)$ applied to a random walker on a $2$-dimensional lattice, as this is a discretized version of the Euclidean space $\mathbb{R}^2$, for different embedding dimensions $m$. In this lattice, each node is labeled by a $2$-dimensional vector $(v_x,v_y)$ ($d=2$) where $v_x,v_y \in [1,1000]$ are natural numbers.
We find that $C_m(r)$ evidences a scaling region with $\beta_m\rightarrow 2$, what suggests that the underlying network has a well defined dimension equal to $2$, {\it i.e.}, the Hausdorff dimension of the plane.

As a further validation we address the case of a fully connected network, in which all the nodes are connected among each other, which is usually seen as the discrete version of an infinite-dimensional space. Note that a fully connected network does not have a natural spatial embedding and therefore, for the sake of simplicity, we label each node by a single real number $v \in [0,1]$ ($d=1$). In the right panel of the Fig. \ref{validation} we represent the correlation sum of the generated trajectory for different embedding dimensions $m$ (bottom panel). In all the cases we find a clear  scaling showing different slopes $\beta_m$. In the top panel of the same figure we plot the estimated value $\beta_m$ as a function of $m$, pointing out a linear dependence $\beta_m\approx m$. This lack of convergence suggests that the underlying structure is infinite-dimensional, as expected.

\begin{figure}
\centering
\includegraphics[width=0.85\columnwidth]{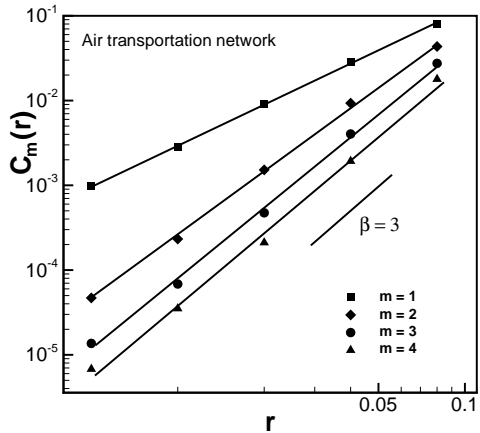}
\caption{Log-log plot of the correlation sum $C_m(r)$ as a function of similarity $r$, for a series of $2\cdot10^4$ data extracted from a random walker of $2\cdot10^4$ steps over the worldwide air transportation network (see the text), for increasing embedding dimensions $m$. The correlation exponent converges to $\beta=3$ (for $m=4$, we obtain an estimate $\beta=2.95\pm0.1$).}
\label{airport}
\end{figure}

\begin{figure}
\centering
\includegraphics[width=1.0\columnwidth]{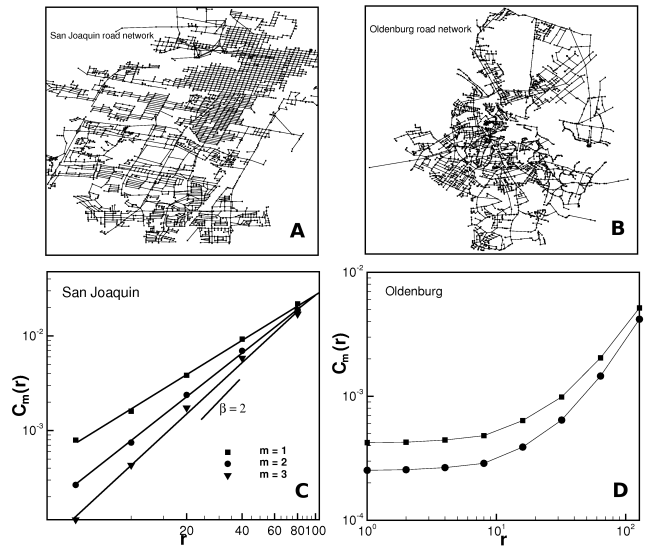}
\caption{(A, B) Samples of San Joaquin (A) and Oldenburg (B) urban networks (see the text). The former is a recently founded city whose structure shows a top-bottom organization and a grid-like aspect, whereas the latter is a city that dates back to the twelve century and shows a self-organized shape without any evident symmetry.   (C, D) Log-log plots of the correlation sum $C_m(r)$ as a function of similarity $r$, for a series of $4\cdot10^4$ data extracted from a random walker of $2\cdot10^4$ steps in the San Joaquin and the Oldenburg urban networks respectively, for increasing embedding dimension $m$. Results suggest that only San Joaquin has a well defined dimension.}
\label{net1}
\end{figure}

Once we have validated the method in synthetic networks we tackle the characterization of real-world networks. We first address the case of the global air-transportation network \cite{guimera}, as this is a paradigmatic spatially embedded network whose dimension has been recently claimed to be larger than two \cite{boxcovering3}. This network is formed by $N=3618$ nodes (the airports) and $L=13514$ links denoting the commercial routes among them. As in the case of the $2$-dimensional lattice we label each node $i$ by a vector $v_i=(x_i,y_i)$ ($d=2$) that determines the normalized geographical coordinates of these airports, where $x_i,y_i \in [0,1]$. In Fig. \ref{airport} we show the results of $C_m(r)$ for a random walk trajectory of $2\cdot10^4$ steps, \emph{i.e.}, an original series of $4\cdot10^4$ data. We find an intermediate regime where a scaling $C_m(r)\sim r^{\beta_m}$ shows up, and $\beta_m\rightarrow\beta\approx 3$ for increasing values of the embedding dimension $m$. This value coincides with the box-covering dimension of the air transportation network, as suggested recently \cite{boxcovering3}, pointing out that, albeit embedded in 2-dimensional space, this network has a larger effective dimensionality. Furthermore, note that the random walk has a length of $2\cdot10^4$ steps, thus revealing that it is possible to derive an accurate value of the network dimension with only a rather small amount of local information.

To round off, we explore the dimension of urban networks \cite{road}, and address two paradigmatic cases of urban development: the case of San Joaquin county (California, US), having $N=18623$ nodes and $L=23874$ edges, and that of Oldenburg (Germany), with $N=6105$ nodes and $L=7035$ edges (see the panels $A$ and $B$ in Fig. \ref{net1} for graphical illustrations). In both networks, each node is characterized by a $2$-dimensional vector (x,y) where $x, y \in [0,10000]$ ($d=2$). Notice that San Joaquin is a recently founded city (1920) whose shape is the result of a planning process and, accordingly, displays a grid-like road structure.  Conversely, Oldenburg (Germany) is an old city whose foundation dates back to the twelfth century and whose road pattern is the result of a self-organized growth. In panel C and D of the same figure, we show their respective correlation sum functions. While the case of San Joaquin (panel C) evidences a scaling regime with a correlation dimension converging to $2$ ($\beta=1.83\pm0.1$ for $m=3$), no scaling is found for the self-organized city of Oldenburg (panel D), suggesting that this latter network does not possess a well defined dimension.
These different behaviors deepen on the recently observed structural differences between cities that have grown according to different evolutionary mechanisms \cite{cardillo1,cardillo2}. In the Supplemental Material \cite{SM} we include additional analysis and estimation of the correlation dimension of other real world examples including technological (Internet at the Autonomous System level \cite{AS} and the Italian power grid \cite{PG}) and road (San Francisco \cite{road} and USA \cite{road2}) networks. Their corresponding analysis yields a well defined and justified correlation dimension.
\smallskip

To conclude, in this work we propose an extension of the Grassberger-Procaccia method to estimate the correlation dimension of a complex network from the analysis of the trajectories of random walkers on top of them. Although the original  method was initially designed as a tool to retrieve the attractor dimension of low-dimensional chaotic dynamics, the presence of temporal correlations in stochastic dynamics (here induced by the geometry of the network) also produces similar behaviors under this celebrated framework  \cite{witten, book}. Thus, in this work we deliberately exploit this property when using random walks as the trajectories under study. This probes the possibility of making use of concepts and tools from the ergodic theory of dynamical systems in the realm of complex networks.


Our results suggest that the dimensionality of spatially embedded networks can be retrieved from this analysis. We highlight that the method only requires local information and it works with rather small time series. This constitutes an advantage for saving memory resources on one hand, and perhaps more importantly, it provides a way to make estimates about the dimension of a network without having global information of its structure. An example of such situation is the routing of information in the Internet, as it is easy to have access to the sequence of IP's a packet navigates through, while having access to the whole Internet map seems unfeasible.

Further work should be done in order {\em (i)} to check in which situations this procedure can be performed, {\em (ii)} to relate the meaning of the exponent found in this work with other exponents recently defined in the network literature, and {\em (iii)} to extend this method to the study of generic networks beyond spatially embedded ones.

\medskip

\acknowledgments
The authors thank F. Papadopoulos and M. Bogu\~{n}\'{a} for providing the dataset of the AS network and S. Meloni for providing the dataset of the Italian power grid. We acknowledge support from the Spanish DGICYT under projects FIS2011-25167, FIS2012-38266-C02-01, FIS2009-13690 and EXPLORA FIS2011-14539-E, by the Comunidad de Arag\'on (Project No. FMI22/10) and Comunidad de Madrid (MODELICO). J.G.G. is supported by MICINN through the Ram\'on y Cajal program. L.L. acknowledges the hospitality of the Department of Condensed Matter Physics at the University of Zaragoza and the Department of Physics at the University of Oxford, where parts of this research were developed.

\begin{table*}[!b]
\begin{ruledtabular}
\begin{tabular}{cccccddd}
Type of Network&Number of Nodes&Size of walker used&Estimated dimension $\beta$\\
\hline
USA road network &$175813$&$5\cdot10^4$&1.5\\
San Francisco Road Network &$174956$&$2.5\cdot 10^4$&2.0\\
Internet (AS) network&$20830$&$2.25\cdot 10^4$&2.5\\
Italian 380kV Power Grid &$310$&$1.25\cdot 10^4$&1.0\\
\end{tabular}
\end{ruledtabular}
\caption{\label{table}Estimated correlation dimension of different real world complex networks. The urban network of San Francisco describes the city in a similar way of a 2D lattice. On the other hand, the entire USA network, interpolates network layers associated to urban cities with network layers associated to roads crossing large inhabited regions. The former is typically (at least for the majority of US cities, which are relatively new and follow from an top-down urban planning) associated to 2D lattice whereas the latter is reminiscent of 1D lattice. The resulting global network interpolates such layers and we find an estimated correlation dimension that converges to $\beta\approx1.5$. The internet map (at the level of autonomous systems) evidences long-distance links and therefore its dimensionality exceeds the dimension of the geographic embedding, converging to $\beta\approx2.5$. Finally, results indicate that the Italian Power-Grid is similar to a 1D lattice with periodic boundary conditions, i.e., a ring.}
\end{table*}
\section{Supplementary Material}
In this supplementary material we complete the results shown in the main text by addressing the analysis  of an additional set of real-world networks. In each case we estimate their correlation dimension and justify its meaning. In particular, we first address two further real-world road networks, namely: (i) the USA road network, and (ii) the San Francisco urban network. We also tackle two real-world technological networks, namely: (iii) the worldwide Internet network at the autonomous systems level, and (iv) the Italian 380KV power grid network. The estimated correlation dimensions for these networks are depicted in table \ref{table}, and the full detailed analysis is outlined below.\\

\noindent \texttt{\underline{USA road network}}\\
\smallskip

The original network \cite{road2} is formed by $175813$ nodes and $179179$ edges, where each node is characterized by a longitude/latitude array (normalized in $[0,10000]$. A geographical embedding of this network is shown in the left panel of figure \ref{net1}, where in the right panel we show a high-precision portion. Note that roads that connect cities evidence a chain graph structure (1D lattice).

\smallskip

\begin{figure}
\centering
\includegraphics[width=0.5\columnwidth]{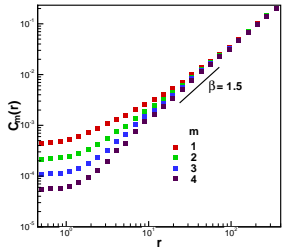}
\caption{Log-Log plot of the correlation sum $C_m(r)$ as a function of similarity $r$, for a series of $\cdot10^5$ data extracted from a random walker of $5\cdot10^4$ steps over the USA road network (see the text), for increasing embedding dimensions $m$. We find evidence
of an increasing scaling regime, with a correlation dimension that converges to $\beta\approx1.5$.}
\label{CM_USA}
\end{figure}

The correlation sum $C_m(r)$, calculated for a series of $10^5$ data extracted from random walker of $5\cdot10^4$ steps is plotted in log-log scales in figure \ref{CM_USA}, for different values of $m$. Note that in every case we find an intermediate regime where a scaling takes place, for all values of $m$. The slope of the scalings saturates to a value $\beta\approx1.5$. Note that if the network evidenced a pure chain structure (1D lattice), its limiting dimension would be $D_2=1$. The estimation seems to interpolate between a chain network and a 2D lattice one. This can be justified as the network is indeed formed by two different subgraphs. The first one is associated to the network structure at the cities and their surroundings: this part of the network is (at least for the majority of US cities, which are relatively new and follow from an top-down urban planning) very much grid-like. On the other hand, the cities are connected by very long roads that cross large large inhabited regions. The network at this level is typically chain-like. Thus, the resulting global network merges these two different network layers, and the correlation dimension efficiently retrieves this merging.\\


\noindent \texttt{\underline{San Francisco Road Network}}

\medskip

\begin{figure}
\centering
\includegraphics[width=0.45\columnwidth]{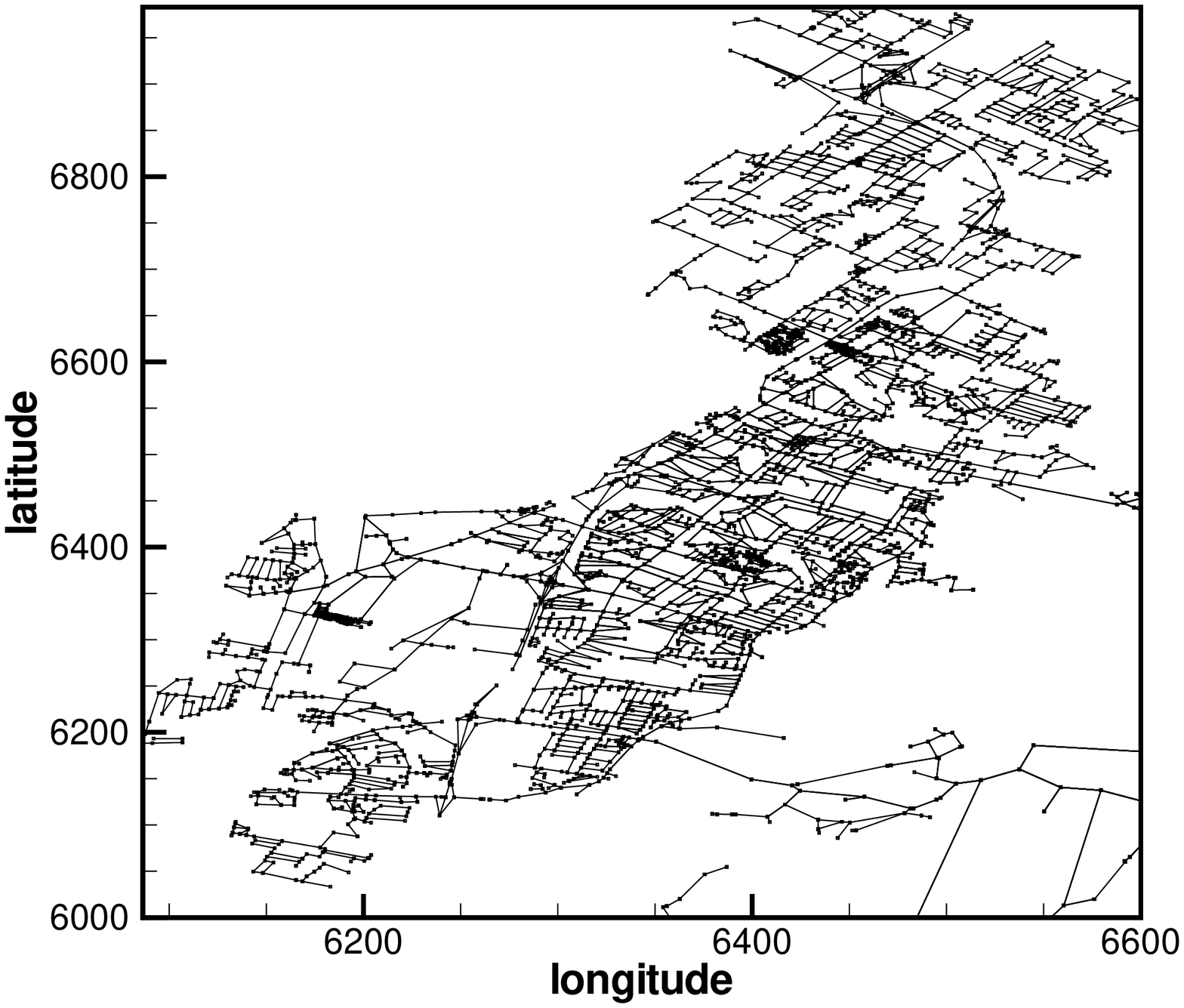}
\includegraphics[width=0.45\columnwidth]{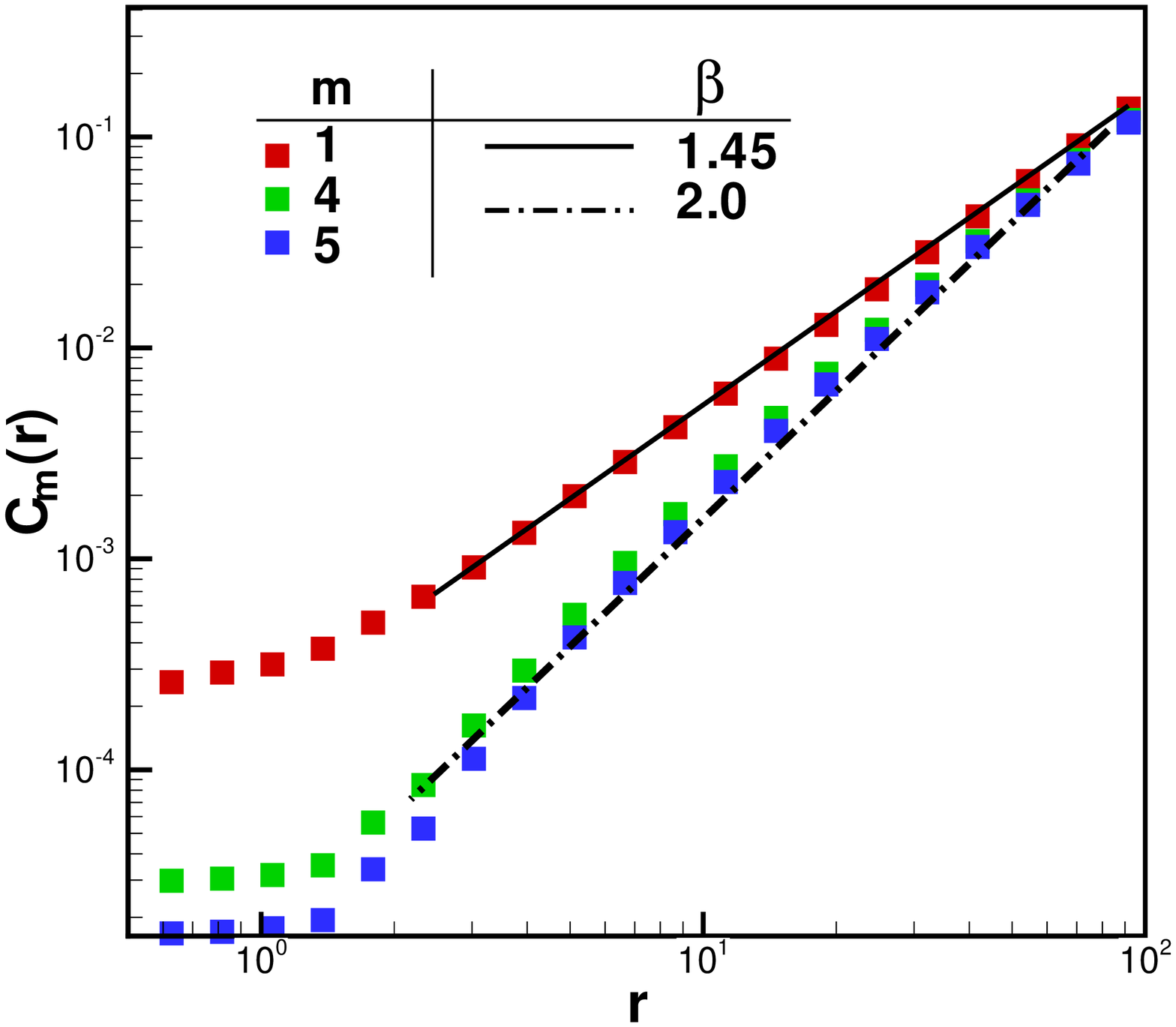}
\caption{(Left panel) Sample of San Francisco road network. Notice that despite common sense, its structure is very much grid-like. (Right panel) Log-log plot of the correlation sum $C_m(r)$ as a function of similarity $r$, for a series of $5\cdot10^4$ data extracted from a random walker of $2.5\cdot10^4$ steps over the San Francisco road network (see the text), for increasing embedding dimensions $m$. We find evidence
of scaling $\forall m$, with an exponent that converges with increasing embedding dimension towards $\beta\approx2.0$, the dimension of a 2D lattice.}
\label{net3}
\end{figure}

The original network \cite{road} is formed by $174956$ nodes. As several nodes work as sources and loads, several nodes correspond to the same geographical location, i.e., the effective network, as regards to the random walk diffusion, is actually much smaller. A geographical embedding of this network is shown in the left panel of figure \ref{net3}.

\smallskip

The correlation sum $C_m(r)$, calculated for a series of $5\cdot10^4$ data extracted from a random walker of $2.5\cdot10^4$ steps diffusing along the network is plotted in log-log scales in the right panel of the same figure, for different values of $m$. Note that in every case we find an intermediate scaling regime for all values of $m$, where the scaling exponent converges to $\beta\approx2.0$, the correlation dimension of a 2D lattice, for increasing embedding dimension $m$. This complies with the grid structure that the San Francisco road network shows.\\

\noindent \texttt{\underline{World wide Internet (Autonomous Systems) network}}\\

\smallskip

\begin{figure}
\centering
\includegraphics[width=0.45\columnwidth]{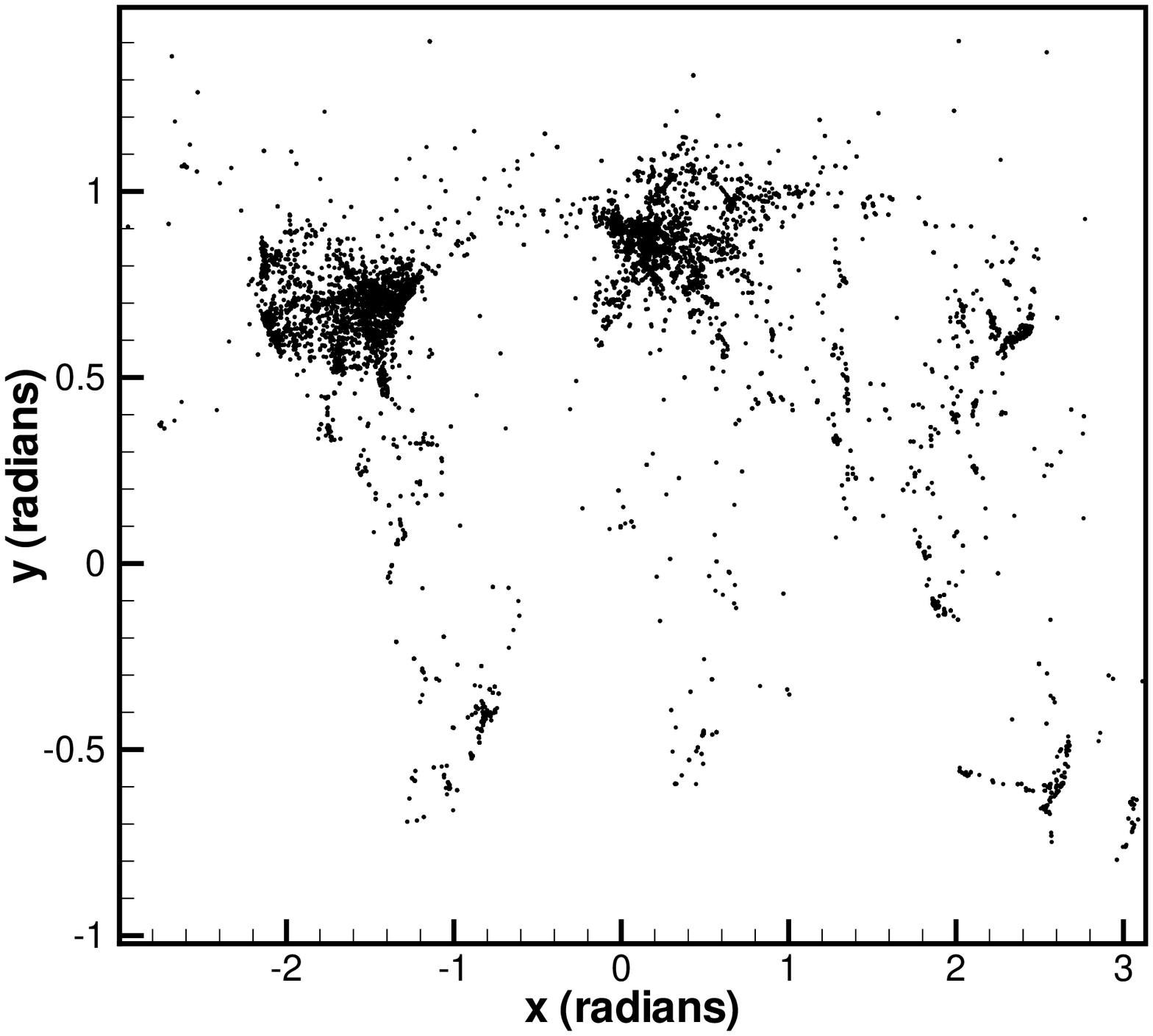}
\includegraphics[width=0.45\columnwidth]{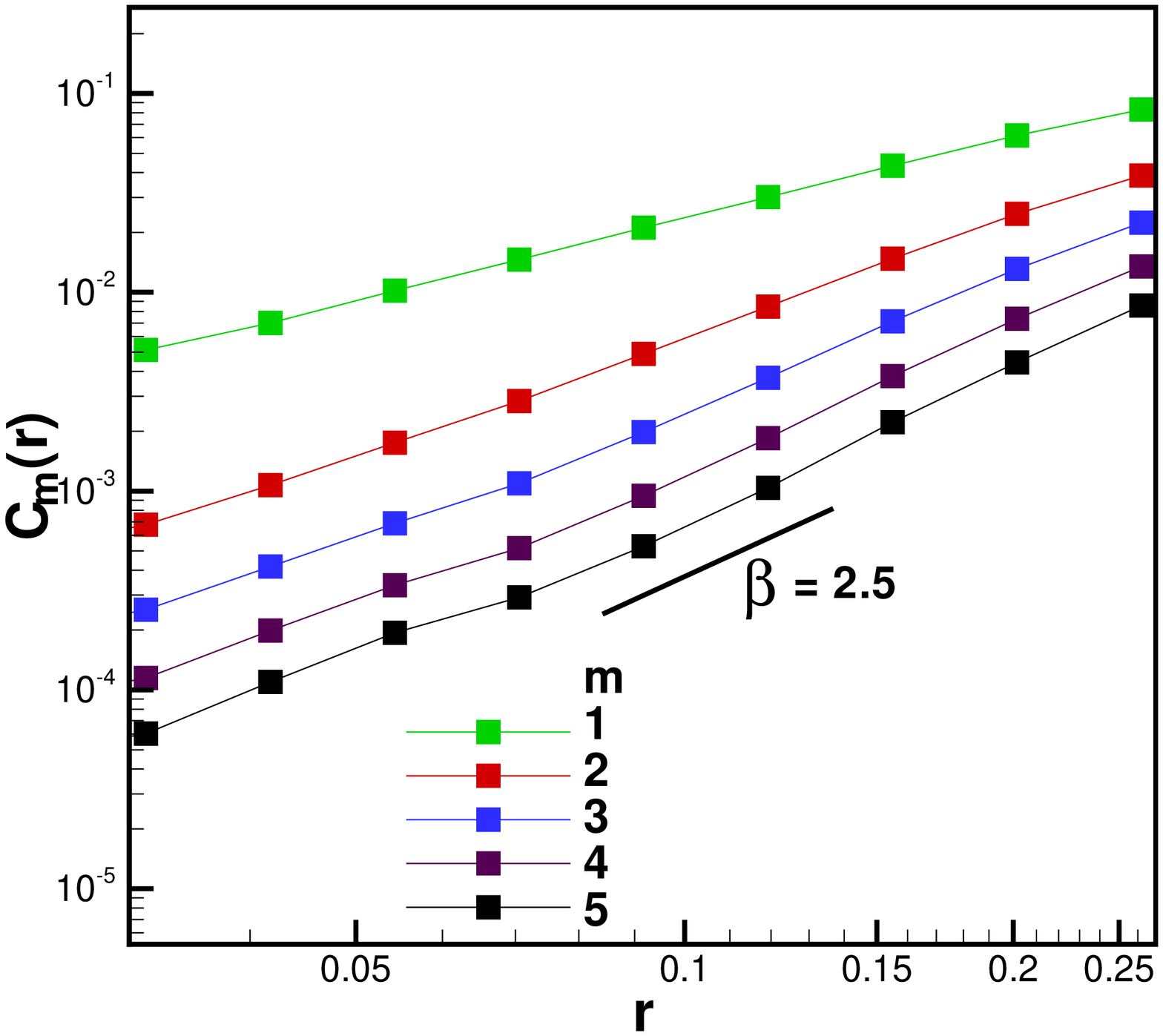}
\caption{(Left panel) Sample of Internet AS network (links not showed). (Right panel) Log-log plot of the correlation sum $C_m(r)$ as a function of similarity $r$, for a series of $5\cdot10^4$ data extracted from a random walker of $2.5\cdot10^4$ steps over the AS network (see the text), for increasing embedding dimensions $m$. We find evidence
of scaling $\forall m$, with an exponent that converges with increasing embedding dimension towards $\beta\approx2.5$.}
\label{net4}
\end{figure}

The original network \cite{AS} is formed by $20830$ nodes. A sample of the geographical embedding of this network is shown in the left panel of figure \ref{net4} (links are not shown in this figure).

\smallskip

The correlation sum $C_m(r)$, calculated for a series of $5\cdot10^4$ data extracted from a random walker of $2.5\cdot10^4$ steps diffusing along the network is plotted in log-log scales in the right panel of the same figure, for different values of $m$. Note that in every case we find an intermediate scaling regime, for all values of $m$, and such scaling converges for increasing values of $m$ towards an exponent $\beta\approx2.5$, which is the estimated correlation dimension of the AS network. This dimension is larger than 2 due to the fact that autonomous systems link worldwide through long-distance physical edges which are Euclidean shortcuts and increase the network's dimension beyond the plane's dimensionality. The dimension is smaller than the one found for the worldwide air transportation system ($\beta\approx3.0$), on agreement with the fact that in that case there exist links (air routes) which connect airports at farther distances.\\

\noindent \texttt{\underline{Italian 380kV Power Grid.}}\\

\smallskip

The original network \cite{PG} is formed by $310$ nodes divided in three categories: sources, where power is inserted (113), loads, where power is extracted (210) and junctions, where power flows across (29). As several nodes work as sources and loads, several nodes correspond to the same geographical location, i.e., the effective network, as regards to the random walk diffusion, is actually much smaller. A geographical embedding of this network is shown in the left panel of figure \ref{net1}.

\smallskip

The correlation sum $C_m(r)$, calculated from a series of $2\cdot10^4$ data was extracted from a random walker of $10^4$ steps diffusing along the network, is plotted in linear-linear scales in the right panel of the same figure, for different values of $m$. Note that in every case we find an intermediate regime where a fairly linear relation takes place, for all values of $m$. This suggests that the correlation estimate already converged for $m=1$ to $\beta\approx1.0$. This can be understood in the following terms: the network studied is very small, and constitutes somehow the kernel of the power grid system. This kernel covers the whole Italy in a circular way, mimicking the shape of a ring network, which is indeed a 1D lattice with periodic boundary conditions.\\

\begin{figure}
\centering
\includegraphics[width=0.5\columnwidth]{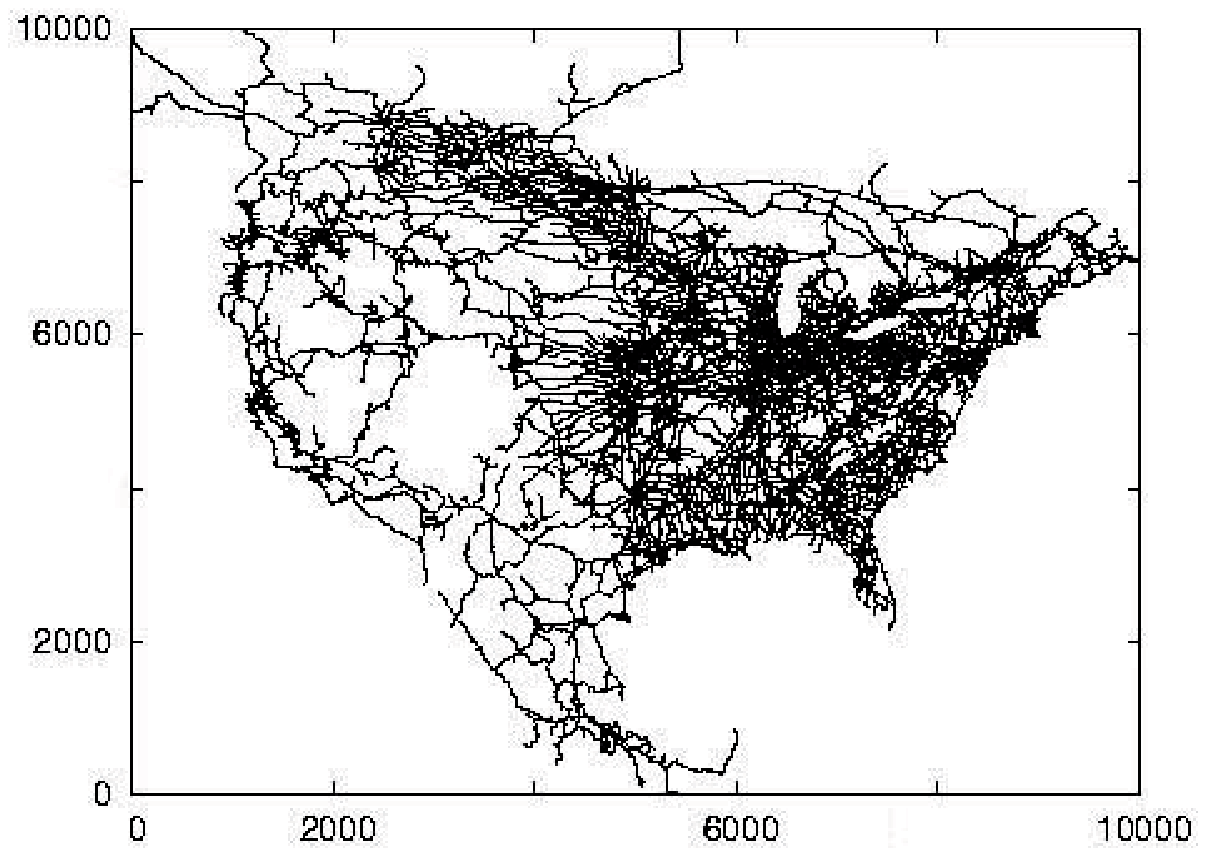}
\includegraphics[width=0.45\columnwidth]{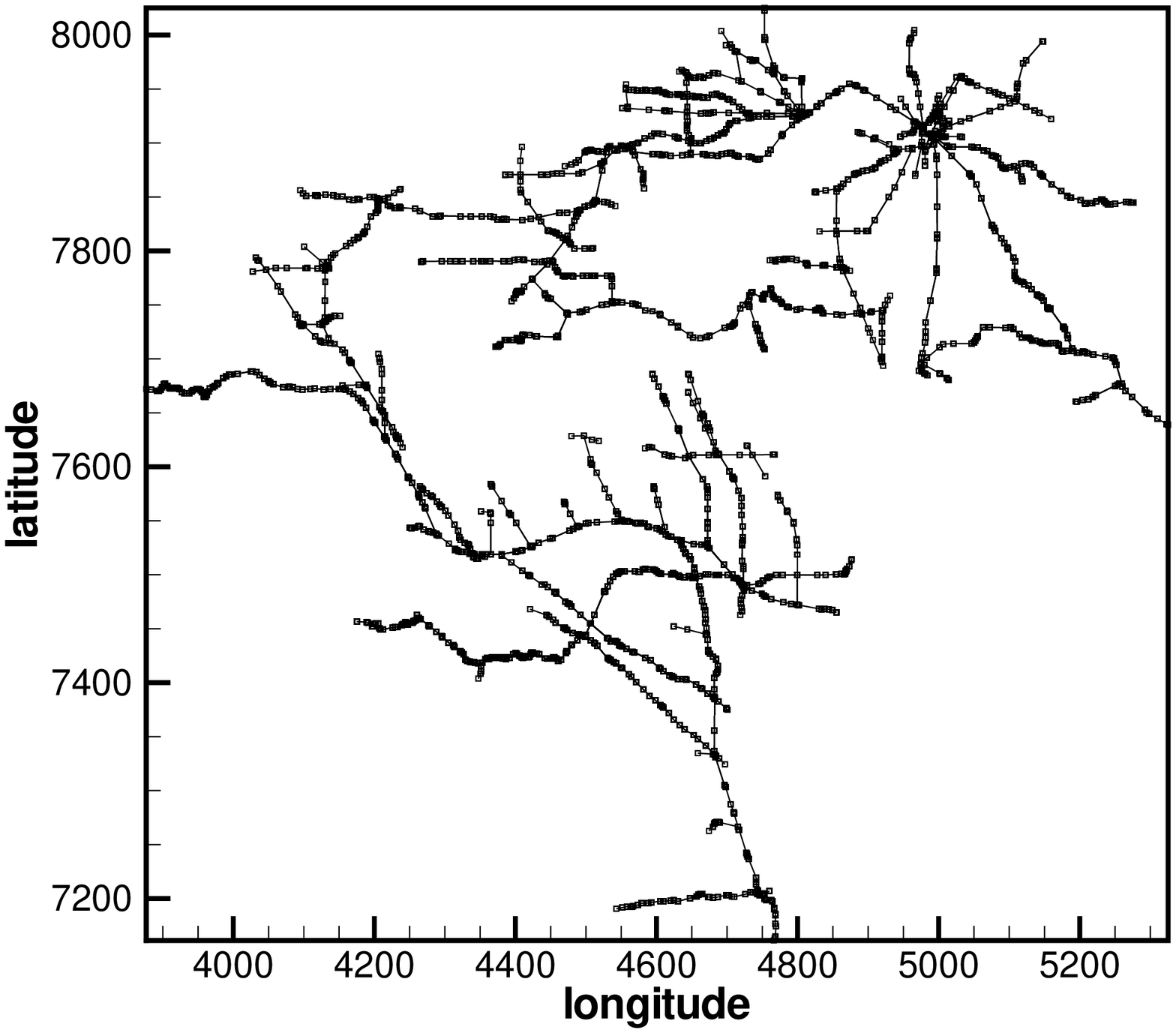}
\caption{(Left panel) USA roads network (see the text). (Right panel) High-precision sample of the same network. Note that many nodes have degree $k=2$, and in this sense the network have many parts which are chain-like. On the other hand, the network covers the entire US country. It therefore interpolates between a 1D and a 2D lattice. Its estimated dimension is on agreement with such interpolation (see figure \ref{CM_USA}). }
\label{net2}
\end{figure}

\begin{figure}
\centering
\includegraphics[width=0.45\columnwidth]{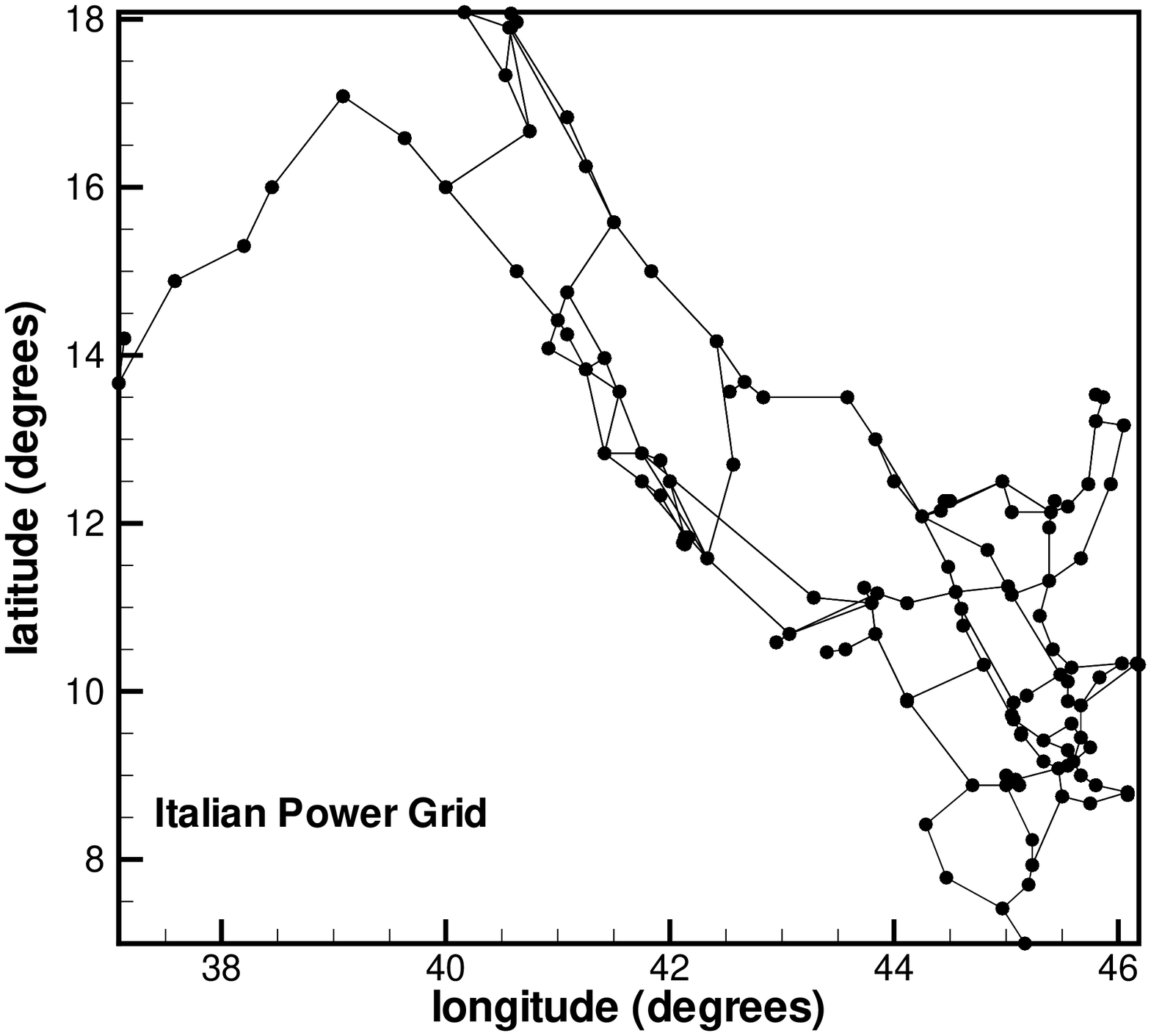}
\includegraphics[width=0.45\columnwidth]{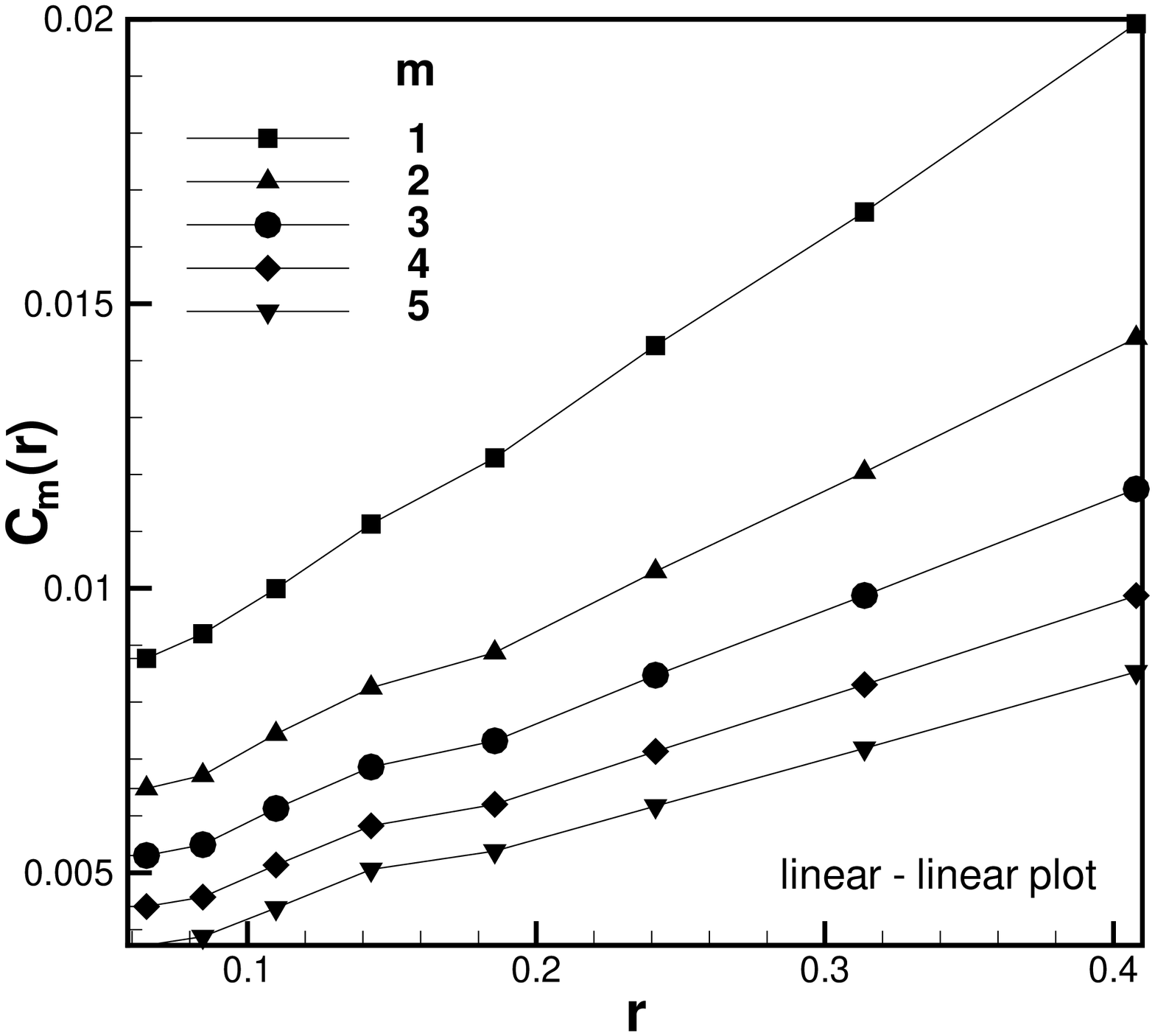}
\caption{(Left panel) Italian 380Kv Power grid. (Right panel) Linear-Linear plot of the correlation sum $C_m(r)$ as a function of similarity $r$, for a series of $2\cdot10^4$ data extracted from a random walker of $\cdot10^4$ steps over the Italian Power Grid (see the text), for increasing embedding dimensions $m$. We find evidence
of linear dependence $\forall m$, suggesting a correlation dimension $\beta\approx1.0$.}
\label{net1}
\end{figure}

\bibliography{apssamp}

\begin{thebibliography}{10}
\bibitem{note1} Note that the trajectories under study are discrete and noise-free series by construction, therefore avoiding some of the classic problems of the original algorithm \cite{sprott}.
\bibitem{note2} The subseries spanned along each space direction are required to have similar mean and variance.
\bibitem{note3} The value of $r$ depends on the amplitude range of the series. As a general recipe, if each of the components of the series $x_i \in [0,x_{max}]$, then $r<x_{max}/100$.
\end{thebibliography}


\end{document}